\documentclass[aps,prl,twocolumn,floatfix,nofootinbib,showpacs]{revtex4}
\usepackage{epsf,amsmath,amssymb,verbatim}
\begin{document}
\newcommand{\be}{\begin{equation}}
\newcommand{\ee}{\end{equation}}
\title{Skeleton and fractal scaling in complex networks}
\author{K.-I. Goh, G. Salvi, B. Kahng, and D. Kim}
\affiliation{\mbox{School of Physics and Center for Theoretical Physics,
Seoul National University, Seoul 151-747, Korea}}
\date{\today}
\begin{abstract}
We find that the fractal scaling in a class of scale-free networks
originates from the underlying tree structure called skeleton,
a special type of spanning tree based on the edge betweenness centrality.
The fractal skeleton has the property of the
critical branching tree. The original fractal networks are viewed
as a fractal skeleton dressed with local shortcuts.
An in-silico model with both the fractal scaling and the scale-invariance
properties is also constructed.
The framework of fractal networks is useful
in understanding the utility and the redundancy in networked systems.
\end{abstract}
\maketitle
The emerging unifying concepts such as the small-world property \cite{ws},
the scale-free behavior \cite{ba}, and the hierarchical
modularity \cite{ravasz} now constitute our basic understanding of
the organization of complex networked systems, which appear in as
diverse examples as the world-wide web, the social networks, and
the biochemical reaction networks inside cells.
The small-world property refers to the one that the average
separation $\langle D \rangle$ between pairs of vertices in the
network scales at most logarithmically in the total number of
vertices $N$ in the system, $\langle D\rangle \sim \ln N$.
The scale-free behavior means the lack of characteristic scales
in the number of links $k$ a vertex has, called the degree,
manifesting itself in the form of a power-law degree distribution
$ p_d(k)\sim k^{-\gamma}$ for large $k$.
Recent discovery of fractal scaling and topological
self-similarity in the world-wide web and
the metabolic networks \cite{ss},
however, raised a new perspective on our view of such networked
systems. The fractal scaling stands for the power-law relation
between the minimum number of boxes $N_B$ needed to cover
the entire network and the size of the boxes $\ell_B$,
\be N_B(\ell_B)\sim \ell_B^{-d_B},\ee
with a finite fractal dimension $d_B$ \cite{feder}.
The self-similarity here
refers to the scale-invariance of the degree distribution
under the coarse-graining with different box sizes (length scales)
as well as under the iterative application of the coarse-graining
(the network renormalization) \cite{ss,bjkim}.
It has been observed, however, that not all networks are
fractal and the most random network models proposed yet are
not fractal, either.
This poses a fundamental question on the origin of the fractal scaling
observed in the real-world networks \cite{ss,strogatz,yook,havlin}.
In this Letter, we show that the fractal property of the
network can be understood from its underlying tree structure.

While highly entangled as a network looks, a more simple structure
is embedded underneath it, that is the spanning tree.
A spanning tree is a tree composed of $N-1$ edges
in a way that they connect all the $N$ vertices in the network.
Of particular significance is the so-called skeleton \cite{skeleton}
of a network. The skeleton is a particular spanning tree,
which is formed by the edges with highest betweenness centralities
\cite{freeman,gn} or loads~\cite{static}.
The remaining edges in the system are called shortcuts, which
contribute to forming loops.
The skeleton of a scale-free network is also scale-free but
with different $\gamma$.
Since the betweenness centrality is related to
the amount of information flow along a given edge,
the skeleton can be considered as the ``communication kernel'' of
the network \cite{skeleton}.
If a network is organized in a modular way,
as it is believed to be so for the world-wide web
and the biological systems, the inter-modular
connections offer communication channels across the modules,
thus gaining high betweenness centralities.
By construction, the skeleton is composed preferentially of such
high-betweenness inter-modular connections, which will preserve
the modular structure while greatly simplifying the complexity.
Furthermore, if the modular structure is distinct enough,
{\it i.e.}, there is a rather clear-cut separation between modules,
we can expect that even a random spanning tree can capture the modular
structure.
Thus by looking at the properties of its spanning trees, we can
visualize more easily the topological organization of the network.

\begin{figure*}
\centerline{\epsfxsize=\linewidth \epsfbox{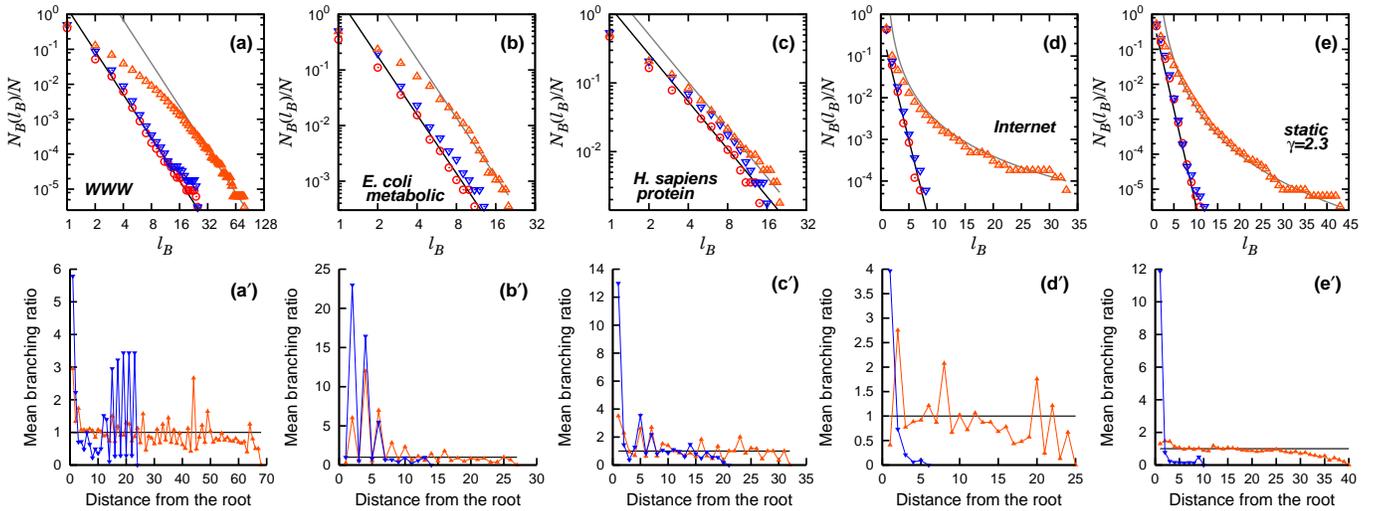}}
\caption{(Color)
{\bf (a--e)} Box counting analysis of original networks ($\circ$, red))
and their skeleton ($\triangledown$, blue) 
and random spanning tree ($\triangle$, orange).
Shown are cases for the world-wide web (a), the metabolic network of
{\it Escherichia coli} (b), and the protein interaction network
of {\it Homo sapiens} (c), the Internet at the autonomous systems
level as of the year 2004 (d), and the static model network with
$\gamma=2.3$ and $\langle k\rangle=4$ (e).
In (a--c), we also show two guidelines for the fractal scaling
of the original network (black) and its random spanning tree (gray).
Slope of each guideline is (a) $-4.1$, (b) $-3.5$,
and (c) $-2.3$. In (d--e), the black line is a fit to the exponential
function and the gray one is to a power law.
{\bf (a$'$--e$'$)} {Branching analysis.}
The mean branching number as a function of distance
from the root for the skeleton ($\triangledown$, blue)
and the random spanning tree ($\triangle$, orange) 
of the networks in (a--e).  For (a$'$--c$'$),
both the skeleton and the random spanning tree fulfill the criticality
condition, $\langle m \rangle=1$ (horizontal line),
as the distance from the root increases,
while for (d$'$--e$'$), the mean branching number of the skeleton
decays to zero with no plateau at $\langle m \rangle=1$.
}
\end{figure*}

With the underlying skeleton and random spanning tree, here we
perform the fractal scaling analysis by measuring $N_B(\ell_B)$
for several real-world networks and network models \cite{method}.
Comparison of the fractal scalings in each original network with
the corresponding spanning trees reveals distinct patterns according to
the presence or the absence of fractality in the network.
For the fractal networks, such as the world-wide
web [Fig.~1(a)] \cite{www}, the metabolic network of
{\it Escherichia coli} [Fig.~1(b)] \cite{metabolic},
and the protein interaction network of {\it Homo sapiens}
[Fig.~1(c)]~\cite{dip}, the numbers of boxes needed to cover
the original network and its skeleton are almost the same.
Moreover, the random spanning tree, while possessing a different
statistics of $N_B$, shows nevertheless the same fractal
dimension $d_B$.
This is surprising, because in the world-wide web, for example,
more than $2N$ edges (shortcuts) are added onto the skeleton
(the average degree of the world-wide web is $6.7$),
which is a tremendous number in
the graph-theoretical sense, and by no means a minute perturbation.
Such a robustness of fractal scaling in the world-wide web shows
that even though the network is far from being a tree,
the shortcuts are distributed in a way that they preserve the
fractality and modularity.
In other words, shortcuts are mainly
present inside modules and the connections between different modules
are largely made through the skeleton.
This topological structure can be measured by
the fraction of intra-branch shortcuts among the total number of
shortcuts, a branch being the subtree connected to the most connected
vertex. We find that the ratio is 0.78, 0.33, and 0.45 for Figs.~1(a),
1(b), and 1(c), respectively. On the other hand, other networks exhibit
different features in the fractal scaling analysis. For example,
the Internet autonomous systems [Fig.~1(d)] or the static model
with $\gamma=2.3$ and $\langle k\rangle=4$ [Fig.~1(e)],
the box counting number of the original network and the skeleton
decays with $\ell_B$ much faster than that of
the random spanning tree, and the fraction of intra-branch shortcuts is
small: 0.087 for Fig.~1(d) and 0.015 for Fig.~1(e). Also for
the social networks, such as the actor network and the collaboration
network, the $N_B(\ell_B)$ curves are appreciably different from those
of the skeletons, implying that the global topology
of the social network is highly interwoven on a large scale
to form a more compact structure \cite{som}.

The scaling behavior of the box counting relation Eq.~(1) in Figs.~1(a)--1(c)
for the original network and its skeleton suggests that the
fractal property of the network originates from that of the skeleton.
In addition, we argue here that the criticality in the topology of
the skeleton is required for a network to be a fractal:
The tree structures such as the skeleton and the random spanning
tree may be seen as generated through a multiplicative branching
process starting from a root vertex~\cite{harris}.
At each branching step, each vertex born in the previous step
generates $m$ offsprings with probability $b_m$.
The criticality condition means the average branching number,
\begin{equation}
\langle m\rangle \equiv\sum_{m=0}^{\infty} m b_m =1.
\end{equation}
Thus the branching tree grows perpetually with offsprings neither
flourishing nor dying out. In this case, when $b_m\sim m^{-\gamma}$,
the number of vertices
$s$ in the tree scales with its linear size $t$ in a power-law
form as $s\sim t^z$ with $z=(\gamma-1)/(\gamma-2)$ for $2 < \gamma
<3$ and $z=2$ for $\gamma >3$~\cite{harris,dslee}, and the tree
structure is fractal with fractal dimension $d_B=z$. Such a
critical branching tree is similar in the topological
characteristics to the homogeneous scale-free tree network
proposed in Ref.~\cite{burda}. To check the validity of our
suggestion, we examine if the criticality condition is fulfilled
for the skeleton and the random spanning trees of the four
real-world networks and the static model in
Figs.~1(a$'$)--1(e$'$). Indeed, for the fractal networks
[Figs.~1(a$'$)--1(c$'$)], both the skeleton and the random
spanning tree fulfill the criticality condition, even though in
reality, the dynamic origins of their formations may well be more
complicated than the pure branching dynamics. In our analysis, the
root is taken as the most connected vertex in the tree. On the
other hand, for non-fractal networks, the mean number of branches
of the skeleton decays to zero rapidly as the distance from the
root increases [see Figs. 1(d$'$) and 1(e$'$)]. A similar behavior
is observed for the actor network as well \cite{som}. Thus the
actor network is not a fractal. However, the random spanning tree
satisfies the criticality condition in all cases, suggesting the
generic fractal structure of this kind of trees as shown in
Ref.~\cite{optimal}. In short, a fractal network contains a
fractal skeleton underneath it, which is perturbed by local
shortcuts, thus preserving its fractal property.

\begin{figure}[!ht]

\begin{minipage}{8.7cm}
\begin{minipage}{0.495\linewidth}
{\epsfxsize=\linewidth \epsfbox{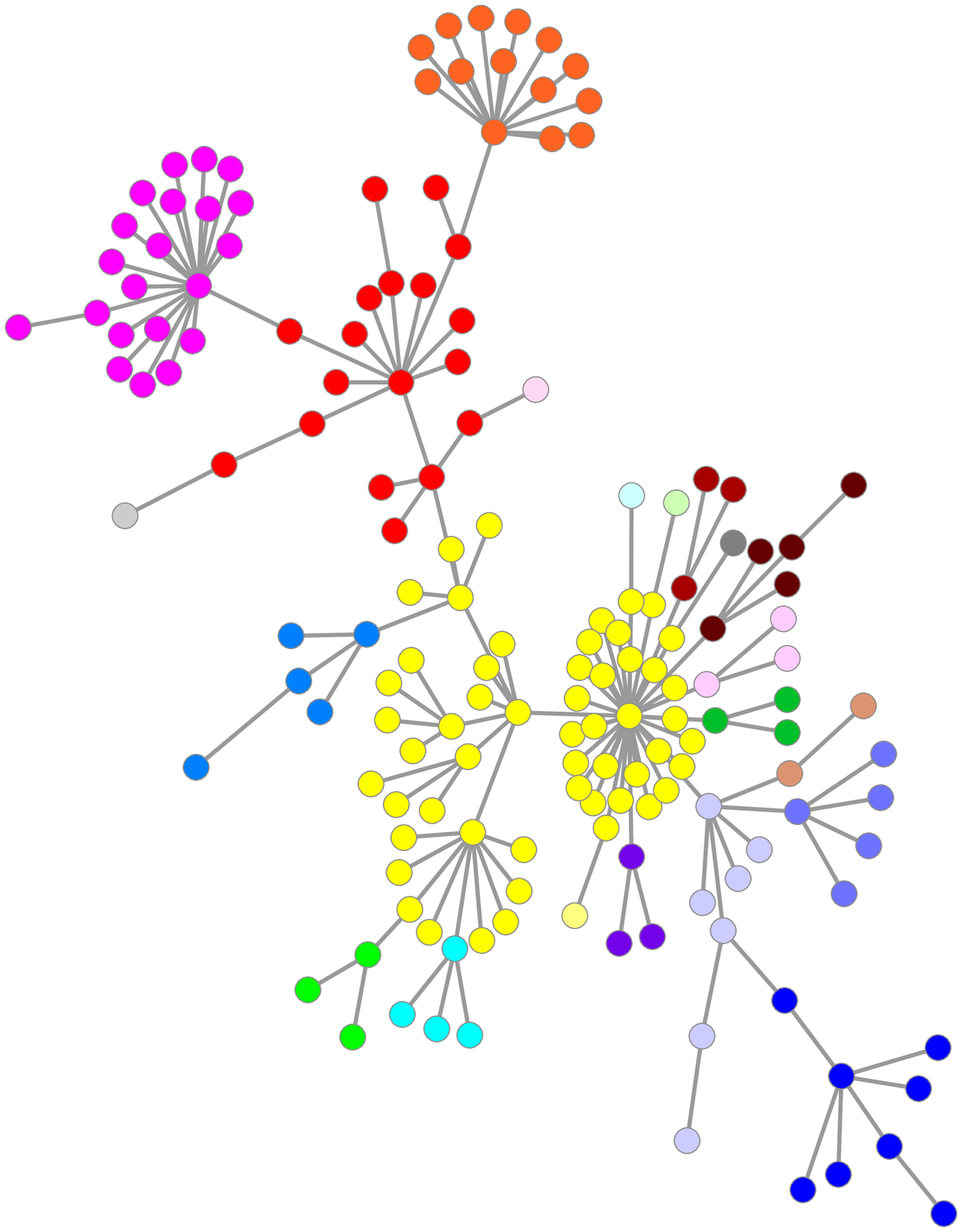}}
\end{minipage}\hfill
\begin{minipage}{0.495\linewidth}
{\epsfxsize=\linewidth \epsfbox{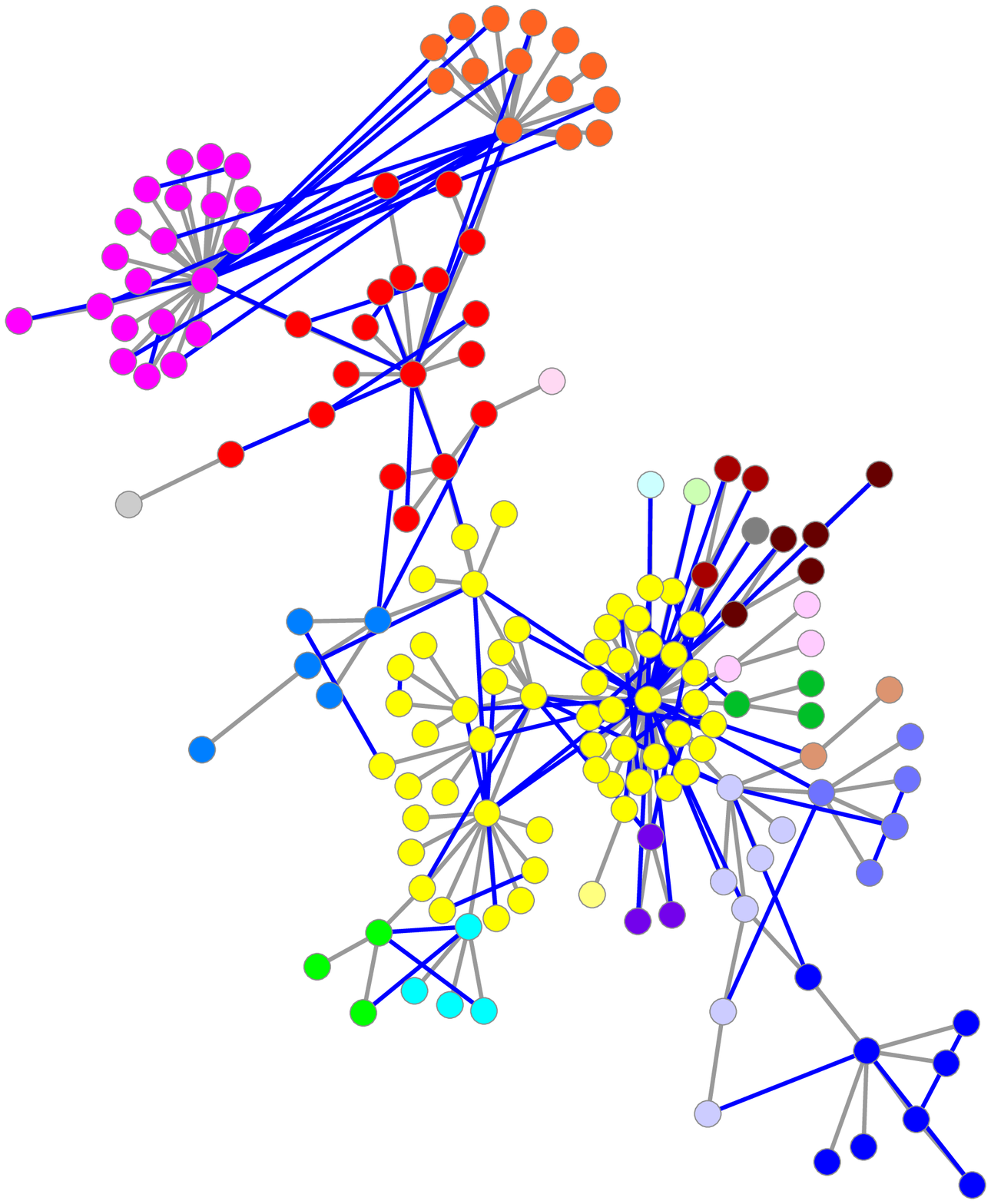}}
\end{minipage}
\begin{minipage}{0.495\linewidth}
{\epsfxsize=\linewidth \epsfbox{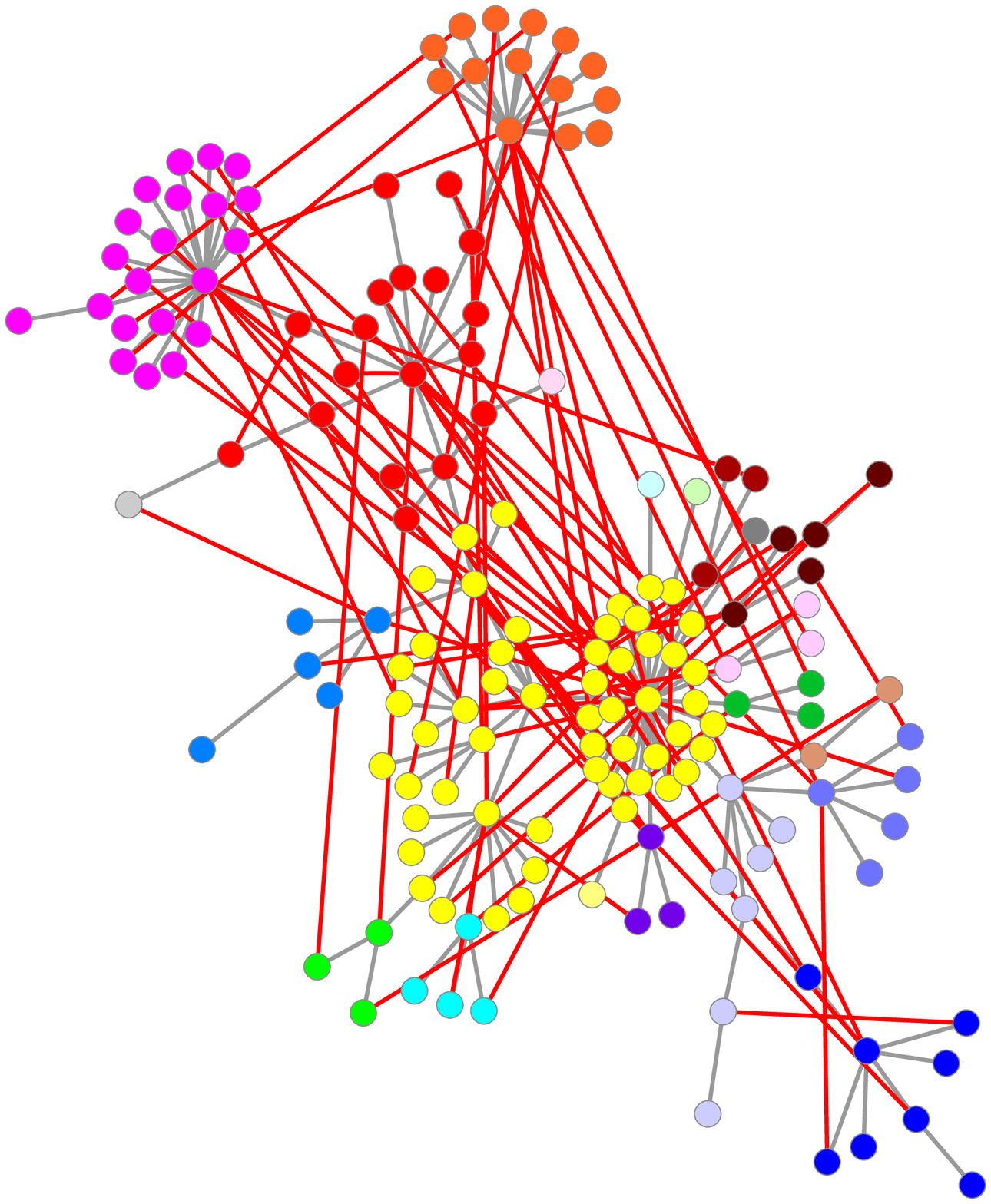}}
\end{minipage}\hfill
\begin{minipage}{0.495\linewidth}
{\epsfxsize=\linewidth \epsfbox{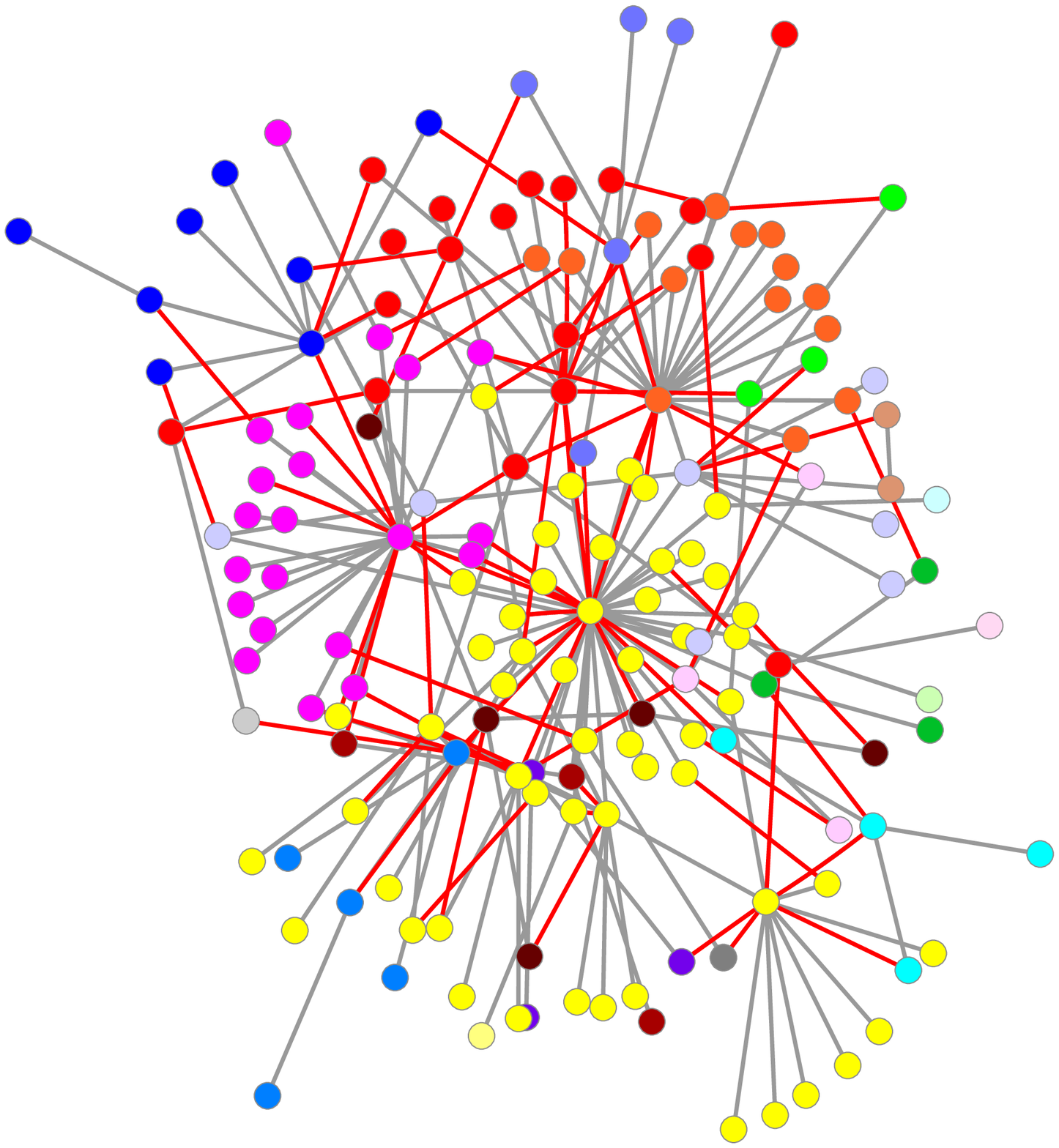}}
\end{minipage}
\begin{minipage}{0.455\linewidth}
\vspace{4mm}
{\epsfxsize=\linewidth \epsfbox{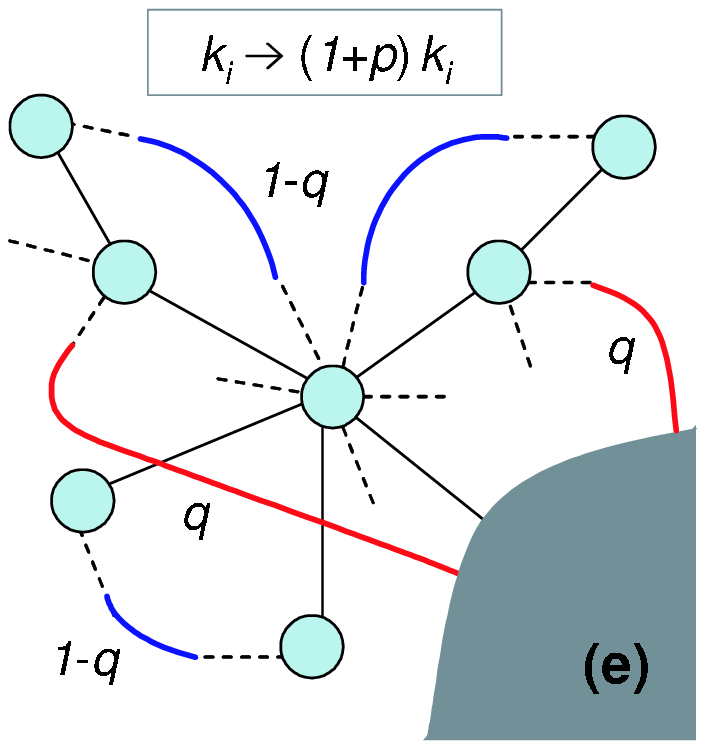}}
\end{minipage}\hfill
\begin{minipage}{0.535\linewidth}
{\epsfxsize=\linewidth \epsfbox{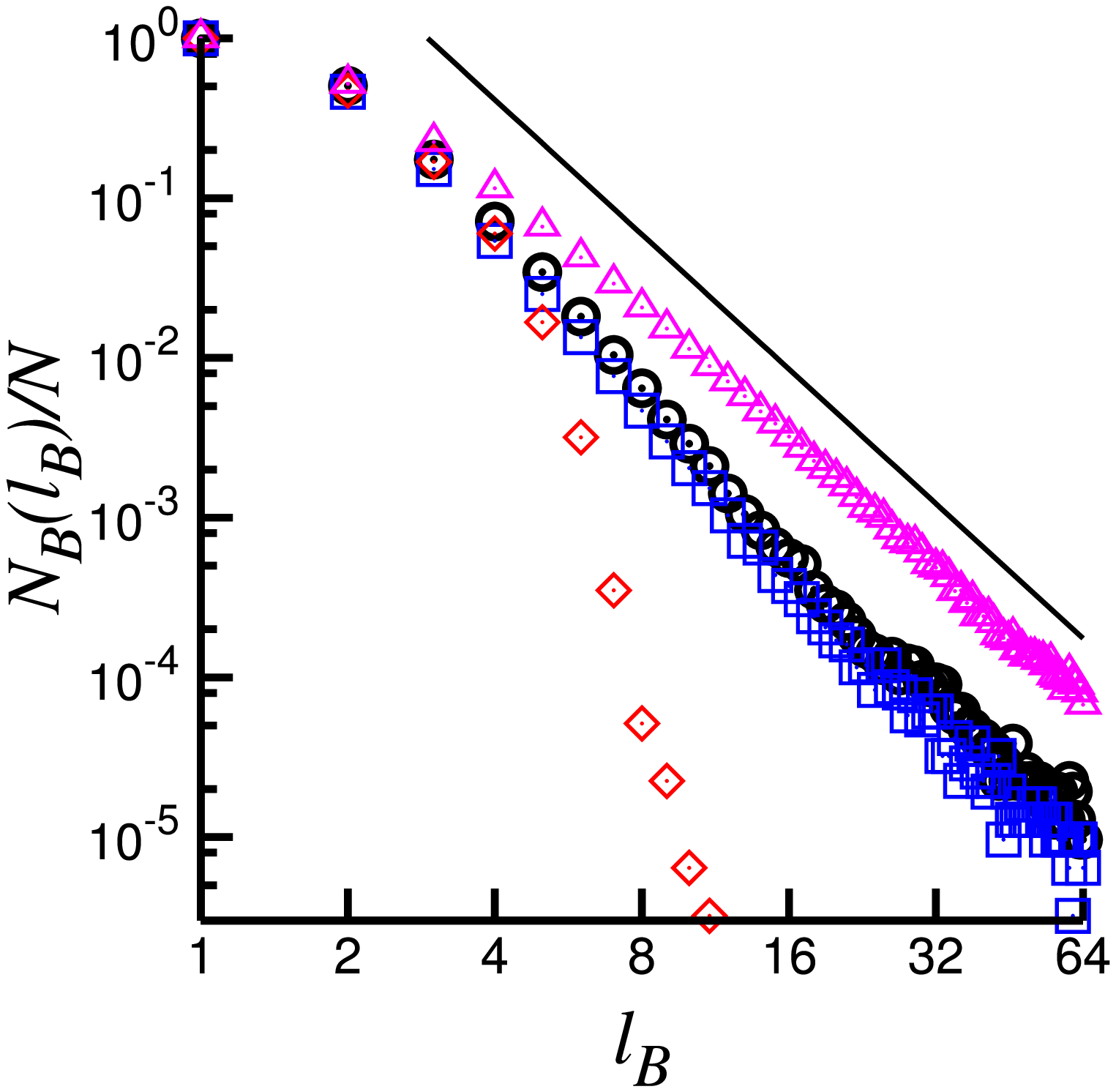}}
\end{minipage}
\end{minipage}
\begin{picture}(200,-1000)(0,0)
\put(80,380){{\bf (a)}}
\put(200,380){{\bf (b)}}
\put(80,235){{\bf (c)}}
\put(200,235){{\bf (d)}}
\put(200,90){{\bf (f)}}
\end{picture}
\caption{
(a) Uncorrelated scale-free tree with the degree
exponent $\gamma=2.3$ and the number of vertices $N=164$.
It is grown by the multiplicative branching rule \cite{harris},
with the branching probability $b_m \sim m^{-\gamma}$.
(b) Fractal model network created by adding local shortcuts 
(green) to the tree in (a).
(c) A non-fractal model network created by adding random shortcuts 
(blue) allowing global connection.
(d) The generic configuration of the network in (c)
generated by Pajek, and the absence of inherent modular structure.
In (b--d), the color of each vertex is that of the corresponding
vertex in (a).
(e) Schematic illustration of the growth rule of the fractal network
model.
The shaded region indicates the rest of the network generated so far.
(f) The fractal scaling analysis of networks with larger size
$N=311,043$, constructed in the same manner as those in (a)-(c). 
Squares (blue) correspond to the fractal network model in (b) 
with $p=0.5$, circles (black) to its skeleton, and the diamonds (red) 
to the non-fractal network model in (c) with $q=1$. 
Solid line is the guideline with slope $\approx-2.8$.
}
\end{figure}

Incorporating all the findings so far, we set up a fractal network
model. The model is based on the multiplicative branching tree
\cite{harris}. We first introduce an exponent $\gamma$ for the
branching probability $b_m \sim m^{-\gamma}$ ($m\ge1$), designed
to produce the desired power-law degree distribution $p_d(k)\sim
k^{-\gamma}$ with $\gamma >2$. The branching probability is
properly normalized to be critical, {\it i.e.}, it satisfies
Eq.~(2). Next, a parameter $p$ is introduced to control the number
of shortcuts added, and hence the mean degree of the network. One
final parameter $q$ accounts for the relative frequency of the
local and the global shortcuts. The construction of
the model network proceeds as follows: {\it i)} A tree is grown by
the multiplicative branching rule with branching probability
$b_m$. {\it ii)} After the branching process, every vertex 
increases its degree by a factor $p$ and attempts to make shortcuts 
to its local neighbors. {\it iii)}
For each successful shortcut in {\it ii)}, with probability $q$,
we replace it by reconnecting it to a randomly chosen vertex, not
restricted to its local neighbors. 
In the latter case, we choose the vertex with a weight in proportion to its
degree in the branching tree, so as to maintain the same power-law
scaling of the degree distribution. This rule is schematized in
Fig.~2(e)~\cite{som}. Fig.~2(a) is an illustrative example of a branching
tree of size $N=164$ and $\gamma={2.3}$. The vertices of the tree
are colored according to which box they belong to in a particular
box-covering with $\ell_B=2$. For
such a tree, even a simple graph drawing algorithm, such as Pajek
\cite{pajek}, can capture its inherent hierarchical structure.
Fig.~2(b) shows the fractal network structure with the addition of
local shortcuts, where we use $p=0.5$ and $q=0$. The
hierarchical modular structure presented in the tree network
[Fig.~2(a)] persists. On the other hand, the network with $q=1$
shown in Fig.~2(c), in which the same number of shortcuts as in
Fig.~2(b) are attached, does not retain the modularity. Such an
absence of modularity can be readily seen in Fig.~2(d), a
different layout of the same network as Fig.~2(c) generated by
Pajek in an unsupervised manner. Consequently, the fractal
property is preserved in the case of $q=0$, whereas it is not for
$q=1$, as is clearly revealed by the box counting analysis in
Fig.~2(f).

It is noteworthy that there exists an important distinction
between the present model and other scale-free trees
such as the Barab\'asi-Albert tree \cite{ba} and the geometrically growing
scale-free tree \cite{jung}. Such models are not fractal,
because they do not fulfill the criticality condition. Indeed,
their mean branching rate decreases to zero monotonically as the
branching proceeds, without a plateau at $\langle m\rangle\approx 1$
\cite{som}. Such a type of tree network was classified as ``causal" trees
in Ref.~\cite{causal}.
Note also that the fractal trees are not small-worlds.
However, with a small number of global shortcuts, e.g., $q=0.01$
in our model with $N\sim3$$\times$$10^5$,
the network turns into a small world. This is seen in the mean
box mass versus $\ell_B$ plot in the cluster growing method 
analysis \cite{som}.

The fractal network model is self-similar.
To check it, we perform the coarse-graining through the box counting method
by replacing each box with a single super-node,
and connecting them if any of their member vertices is connected \cite{ss}.
We find that the degree distribution of the fractal network model
is invariant
under the coarse-graining by the boxes with different sizes.
In addition, the degree distribution is invariant under successive
applications of the coarse-graining transformation \cite{som}.
It is interesting to note that some networks are self-similar,
that is, exhibit the scale-invariant degree distribution, yet
are not fractal. Typical example of such networks is the
Internet \cite{som}.
So the fractality and the self-similarity do not always imply each other
in complex networks.

The framework of fractal network is helpful to understand, for
example, the utility and the redundancy in the metabolic networks
from a purely topological aspect. The high flux backbone in the
metabolic network of {\it E. coli} obtained through the flux
balance analysis was shown to be composed of many branches with
few inter-branch connections and to merge into the biomass
reaction \cite{meta}. Obviously, its topological shape resembles
the branching tree skeleton rooted from a vertex with the largest
number of connections if the direction in edge is ignored. On the
other hand, recent {\it in silico} flux analysis~\cite{ghim} has
shown that the metabolic network of {\it E. coli} contains high
density of backup reactions (redundancies) for a given condition.
Such a reaction is barely used in the normal condition, but takes
up a high flux when a certain reaction on the backbone is blocked.
When the simultaneous blockade of such a reaction pair blocks the
biomass production, they are called synthetic
lethal~\cite{epistasis}. Most synthetic lethal reactions are
located very close to each other, being apart in three reaction
steps or less along the metabolic network. Also they are mostly in
the same functional category. Thus the reactions with high (low)
flux in wild type can be regarded as the edges on the skeleton
(shortcuts) in the framework of the fractal network.

The critical branching tree that can be found in various phenomena
such as earthquake processes, population and biological dynamics,
epidemics, social cascades, {\it etc.}, also appears in the
skeleton of fractal networks, as we found here. While the evolving
process of the fractal networks would be complex and diverse
depending on the specific systems, the underlying structure, the
skeleton, has the topology of the critical branching tree.
Identifying such a simple structure underneath is a step forward
towards further studies on the renormalization and the
universality in complex networks.

This work is supported by the KRF Grant No. R14-2002-059-01000-0
in the ABRL program funded by the Korean government MOEHRD.
G.S. gratefully acknowledges financial support from the Swiss National
Science Foundation under the fellowship PBEL2-106182.

%%%%%%%%%%%%%%%%%%%%%%%%%%%%%%%%%%%%%%%%%%%%%%%%%%%%%%%%%%%%%%%%%%%%%%%%%%%%%
% Bibliography
%%%%%%%%%%%%%%%%%%%%%%%%%%%%%%%%%%%%%%%%%%%%%%%%%%%%%%%%%%%%%%%%%%%%%%%%%%%%%


\begin{thebibliography}{99}
\bibitem{method} We use a slightly different method of box counting, the burning method, that gives the same result as in \cite{ss} but is easier to implement.
\end{thebibliography}
\end{document}